\documentclass[aps,prl,twocolumn,showpacs,groupedaddress]{revtex4}
\usepackage{graphicx}  
\usepackage{dcolumn}   
\usepackage{bm}        
\usepackage{amssymb}   
\usepackage{verbatim}

\usepackage{amsmath}
\usepackage{amsfonts}

\begin{document}

\title{Energy Decay in Josephson Qubits from Non-equilibrium Quasiparticles}
\author{John M. Martinis$^1$}
\author{M. Ansmann$^1$}
\author{J. Aumentado$^2$}

\affiliation{$^{1}$Department of Physics, University of California,
Santa Barbara, California 93106, USA}

\affiliation{$^{2}$National Institute of Standards and Technology, Boulder Co 80305}

\date{\today}

\begin{abstract}
We calculate the energy decay rate of Josephson qubits and superconducting resonators from non-equilibrium quasiparticles.  The decay rates from  experiments are shown to be consistent with predictions based on a prior measurement of the quasiparticle density $n_\textrm{qp}=10\,/\mu\textrm{m}^3$, which suggests that non-equilibrium quasiparticles are an important decoherence source for Josephson qubits.  Calculations of the energy-decay and diffusion of quasiparticles also indicate that prior engineered gap and trap structures, which reduce the density of quasiparticles, should be redesigned to improve their efficacy.  This model  also explains a striking feature in Josephson qubits and resonators - a small \textit{reduction} in decay rate with increasing temperature.
\end{abstract}

\pacs{03.65.Yz, 74.50+r, 74.25.Nf}
\maketitle

Superconducting integrated circuits are a leading candidate for scalable quantum information processing\cite{generalscqubit}, with recent experiments showing accurate state control of coupled circuits\cite{recentcouple}.  Josephson qubits are relatively straightforward to couple because their relatively large size enables quantum interactions via simple wiring.  However, these same wires also make the qubit susceptible to various decoherence mechanisms, and thus a detailed understanding of all loss mechanisms is critically needed\cite{decoherence}.  Non-equilibrium quasiparticles are known to be an important decoherence mechanism for charge qubits, since quasiparticles hopping on and off the qubit island produce an unpredictable change in state\cite{Joyez1994, Lutchyn2006}.  How do quasiparticles affect the coherence of charge-insensitive qubits, such as phase\cite{Lang2003}, flux, and transmon\cite{Koch2007} devices?

In this letter, we present a simple and general model that describes how non-equilibrium quasiparticles produce energy decay in superconducting qubits and resonators.  We relate predictions to the quasiparticle density $n_\textrm{qp}$\,: based on its prior measurement\cite{Shaw2008}, we calculate an energy decay rate $\Gamma_1$ that could possibly explain current limits.  Although other decay mechanisms, such as surface dielectric loss\cite{TLSloss}, are important, experimentalists need a detailed theory of all mechanisms in order to properly design experiments that separate out and measure energy decay, and to ultimately improve qubit performance.  We conclude that quasiparticles must be considered in present experiments.

Our model is described in four parts.  First, we calculate the energy decay rate for a Josephson qubit, using a generalization of the environmental $P(E)$ theory\cite{PEreview}.  Since the prediction is based on tunneling of non-equilibrium quasiparticles, we next calculate their density versus energy from standard electron-phonon rates, showing the injection energy is unimportant.  Third, we estimate quasiparticle diffusion lengths and comment on the design of effective quasiparticle traps.  Finally, we consider and discuss possible sources of non-equilibrium quasiparticles, using energy estimates.

No equilibrium quasiparticles are expected in a superconductor well below the critical temperature $T \ll T_c$, because of the exponential suppression of excitations from the superconducting gap $\Delta$.  However, an experiment on Cooper-pair boxes has measured a significant quasiparticle density $n_\textrm{qp} \sim 10\,/\mu\textrm{m}^3$, arising from an unknown source\cite{Shaw2008}.  Once excited, quasiparticles can readily tunnel through a Josephson junction.

Charge associated with quasiparticles allows a tunneling event to couple energy between the quasiparticle and the qubit, as described by the environmental $P(E)$ theory\cite{PEreview}.  When a qubit state of energy $E_{10}$ decays,  its energy is added to the final quasiparticle state.  Energy is transferred, however, in only a fraction of the tunneling events for the case of large junction capacitance $C$ that we consider here.  For the simple case of the transfer of charge $q$, the probability for qubit decay from a single tunneling event is $p\simeq (q^2/2C)/E_{10}$.  We extend the  $P(E)$ theory to account for quasiparticle tunneling having both electron-like ($q=-e$) and hole-like ($q=e$) charge transfer, which gives a qubit decay probability\cite{noteHO, supplement}
\begin{align}
P \simeq \frac{e^2/2C}{E_{10}} (u_l u_r + v_l v_r)^2 \ ,
\end{align}
where $u$ and $v$ are the (energy dependent) BCS amplitude factors for the left ($l$) and right ($r$) junction electrodes.

For a tunnel junction with normal state resistance $R_T$, the total qubit decay rate is obtained by summing over all possible quasiparticles and accounting for filled and empty states.  For the tunneling of quasiparticles from left to right, we integrate over all possible initial energies $E$ to obtain a rate
\begin{align}
\overrightarrow{\Gamma}_1= &  \frac{4}{R_T e^2} \int_\Delta^\infty P \rho_l\rho_r f_l[1-f_r] dE \\
= & \frac{1}{R_T C} \int_\Delta^\infty \frac{dE}{E_{10}}
\frac{E(E+E_{10})+\Delta^2}
{\sqrt{E^2-\Delta^2}\sqrt{(E+E_{10})^2-\Delta^2}} \nonumber \\
& \ \ \ \ \ \ \ \ \times f_l (E)[1-f_r (E+E_{10})] \label{eqGamma1} \ ,
\end{align}
where $\rho_{l,r}(E)=E/(E^2-\Delta^2)^{1/2}$ is the normalized quasiparticle density of states, and $f_{l,r}$ is the non-equilibrium occupation probability.  The factors $u_r$, $v_r$, $\rho_r$ and $f_r$ are computed for the final quasiparticle state with energy $E+E_{10}$, and the coherence factor is opposite in sign as for charging effects\cite{Shaw2008}.  The total decay rate $\Gamma_1$ is obtained by also summing the quasiparticle tunneling rate from right to left $\overleftarrow{\Gamma}_1$.  We note the similarity of this formula to the Mattis-Bardeen theory for metallic conductors \cite{Mattis, noteequiv}.

The total density of non-equilibrium quasiparticles is given by $n_\textrm{qp}= 2D(E_F)\int \rho(E) f(E)\,dE$, where $D(E_F)$ is the density of electron states at the Fermi energy.  The integral for the decay rate can be solved for the physically relevant case of contributions from $f(E)$ only near the gap, as motivated below.  For small $n_\textrm{qp}$, the total decay rate\cite{LutchynPhD} is
\begin{align}
\Gamma_1 \simeq \frac{\sqrt{2}}{R_T C}  \Big( \frac{\Delta}{E_{10}} \Big) ^{3/2} \frac{n_\textrm{qp}}{D(E_F) \Delta } \label{eqGammaApx}
\ .
\end{align}
The decay rate scales as the fractional quasiparticle occupation, which is normalized by the total density of Cooper pairs $D(E_F)\Delta$.

We can immediately check the relevance of this prediction by plugging in typical parameters for qubits.  In Ref.\,\cite{Shaw2008}, a quasiparticle density of $10\,/\mu\textrm{m}^3$ was experimentally measured in charge qubits.  For the aluminum superconductor that is typically used, we have $\Delta/k = 2.1\,\textrm{K}$ and $D(E_F)\Delta = 2.8\cdot 10^6/\mu\textrm{m}^3$.  Typical phase qubit parameters are $E_{10}/h = 6\,\textrm{GHz}$, $R_T = 200\,\mathrm{\Omega}$, and $C=1\,\textrm{pF}$, which yields from Eq.\,(\ref{eqGammaApx}) a qubit decay rate $\Gamma_1=1/(2.1\,\mu\textrm{s})$.  This rate is reasonably close to what is measured for our qubit\cite{UCSBphaseT1} $T_1 \simeq 550\,\textrm{ns}$, especially considering Ref.\,\cite{Shaw2008} possibly had lower $n_\textrm{qp}$ because of quasiparticle traps.

Non-equilibrium quasiparticles also produce energy decay in resonators from loss in superconducting wires.  Following the calculations in Ref.\,\cite{GaoThesis}, the resonator quality factor $Q$ can be expressed as a function of the quasiparticle density \cite{supplement}.  Here, the bulk conductivity of the superconductor is given by the Mattis-Bardeen theory, from which the surface impedance can be expressed in various limits.  Geometric factors expressing the non-uniform current density in the coplanar transmission line are also included, to yield in the thick-film limit
\begin{align}
\frac{1}{Q} = \frac{\lambda}{s}\  \frac{g}{g_m}
\gamma \frac{\sqrt{2}}{\pi} \Big( \frac{\Delta}{\hbar\omega} \Big) ^{1/2} \frac{n_\textrm{qp}}{D(E_F) \Delta } \ .
\end{align}
Here, $\lambda \simeq (\pi\mu_0\sigma_n\Delta/\hbar)^{-1/2}$ is the penetration depth corresponding to the surface kinetic inductance $\mu_0\lambda$, $\sigma_n$ is the normal state conductivity, $s$ is the width of the center conductor, $g = 1.0$ and $g_m=0.31$ are geometrical factors expressing the effect of non-uniform current density on the kinetic and magnetic inductance, respectively, and $\gamma = 1/2$ is a factor concerning the interface.  Using $s = 3\,\mu\textrm{m}$, $\lambda= 50\,\textrm{nm}$ for Al, a resonator frequency of 6\,GHz, and quasiparticle parameters described previously, we compute $Q \simeq 10^7$, which corresponds to a decay time of $300\,\mu\textrm{s}$.  This time is a significantly longer than found above for qubits, and agrees with experimental observation that decay times are longer for resonators \cite{bestresonator}.

A more detailed understanding of non-equilibrium quasiparticles can be obtained by calculating their occupation probability $f(E)$ using electron-phonon scattering rates\cite{Kaplan1976}.  We first consider a bulk superconductor.   The lifetime of a quasiparticle at energy $\epsilon$ to scatter to any energy $\epsilon'$ by emitting or absorbing a phonon of energy $\epsilon-\epsilon'$ is
\begin{align}
\Gamma_{\epsilon\rightarrow\epsilon'}^s = &\frac{1}{\tau_0} \int_\Delta^\infty
d\epsilon' \frac{(\epsilon-\epsilon')^2}{(kT_c)^3}
\frac{\epsilon'}{\sqrt{\epsilon'^2-\Delta^2}}
(1-\frac{\Delta^2}{\epsilon\epsilon'}) \nonumber \\
& \ \ \ \times \frac{1-f(\epsilon')}{|\exp[-(\epsilon-\epsilon')/kT_p]-1|} \ ,
\label{gammas}
\end{align}
where $\tau_0 \approx 400\,\textrm{ns}$ is the measured characteristic electron-phonon time for Al.  The lifetime of a quasiparticle state with energy $\epsilon$ to recombine with another quasiparticle of any energy $\epsilon'$ is
\begin{align}
\Gamma_{\epsilon,\epsilon'}^r = &\frac{1}{\tau_0} \int_\Delta^\infty
d\epsilon' \frac{(\epsilon+\epsilon')^2}{(kT_c)^3}
\frac{\epsilon'f(\epsilon')}{\sqrt{\epsilon'^2-\Delta^2}}
(1+\frac{\Delta^2}{\epsilon\epsilon'}) \nonumber \\
& \ \ \ \times \frac{1}{|\exp[-(\epsilon+\epsilon')/kT_p]-1|} \ .
\label{gammar}
\end{align}
Note the similarity in this formula to Eq.\,(\ref{gammas}), but here the energy of the emitted phonon is greater than $2\Delta$.  With the term  $f(\epsilon')$, this rate is proportional to the density of quasiparticles, implying that the recombination rate slows down once the density of quasiparticles becomes small.

Equations (\ref{gammas}) and (\ref{gammar}) are used to solve for the steady-state occupation probability $f(E)$ of quasiparticles for a given (non-equilibrium) injection rate $r_\textrm{qp}$.  With the recombination rate being proportional to the square of the density of quasiparticles, we solve this nonlinear differential equation numerically.  Because the quasiparticle relaxation rate at high energy is dominated by electron-phonon scattering, which is number conserving for quasiparticles, the exact energies at which the quasiparticles are injected are unimportant for the computed low-energy distribution.  For simplicity, we inject quasiparticles with energies between $2.8\,\Delta$ and $3\,\Delta$.

\begin{figure}[t]
\includegraphics[width=0.47\textwidth,clip=True]{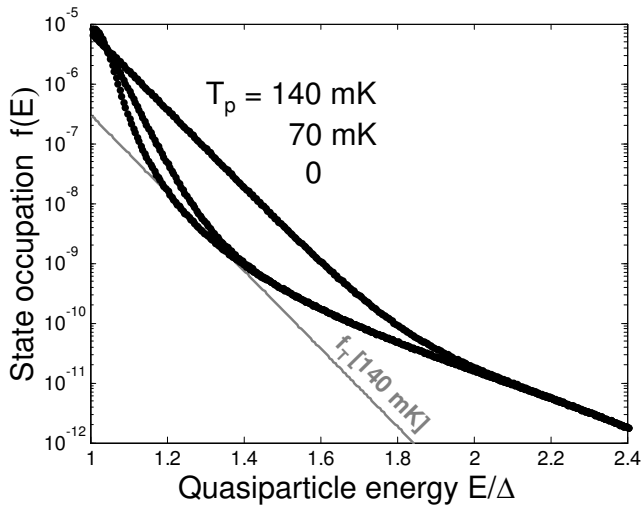}
\caption{Plot of quasiparticle occupation versus energy for 3 phonon temperatures $T_p = $ 140, 70, and 0 mK (top to bottom).  An injection rate was chosen to match the experimental quasiparticle density $10\,/\mu\textrm{m}^3$, equivalent to $n_\textrm{qp}/2D(E_F)\Delta = \int f \rho\, dE = 1.8\cdot10^{-6}$.  The state occupation at energies shown does not depend on the injection energies. The gray line shows the thermal occupation $f_T$ at $140\,\textrm{mK}$. }
\label{FigfE}
\end{figure}

We plot the quasiparticle occupation probability versus energy in Fig.\, \ref{FigfE} for an injection rate $r_\textrm{qp}/2D(E_F)\Delta = 1.7\cdot10^{-10}/\tau_0$ that gives a total quasiparticle density $n_\textrm{qp}/2D(E_F)\Delta = 1.8\cdot10^{-6}$, approximately the same as for Ref.\,\cite{Shaw2008}.  At phonon temperatures $T \gtrsim 170\,\textrm{mK}$, a thermal occupation $f_T$ of quasiparticles dominates.  At $T \sim 140\,\textrm{mK}$, an exponentially decaying occupation is calculated, but at an elevated occupation due to the non-equilibrium quasiparticles.  For temperatures below about $70\,\textrm{mK}$, the curves become independent of temperature.  This calculation is consistent with experiments on Cooper-pair-box experiments\cite{Shaw2008,cpbTdepend}, where non-equilibrium quasiparticle tunneling rates became independent of refrigerator temperature below about $70\,\textrm{mK}$.

The total quasiparticle density at $T=0$ is numerically found to be  $n_\textrm{qp}/2D(E_F)\Delta = 0.14\,(\tau_0 r_\textrm{qp} /2D(E_F)\Delta)^{1/2}$.  We find the computed decay rate is well approximated by Eq.\,(\ref{eqGammaApx}).

\begin{figure}[b]
\includegraphics[width=0.47\textwidth,clip=True]{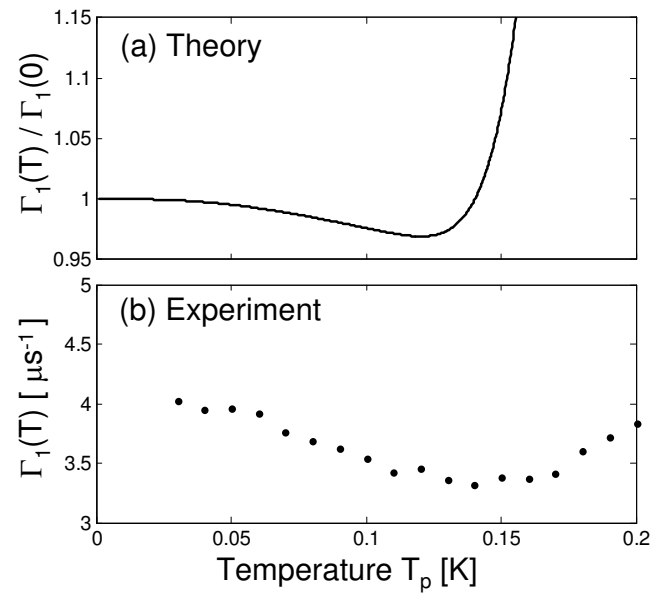}
\caption{(a) Plot of calculated qubit energy relaxation rate versus phonon temperature $T_p$ for a quasiparticle density of $n_\textrm{qp}/2D(E_F)\Delta = 1.8\cdot10^{-6}$ and $E_{10}/h = 6\,\textrm{GHz}$.  This shape of the curve is relatively insensitive to changes in $E_{10}$ and $n_\textrm{qp}$.  Note the small \textit{decrease} in decay rate with increasing temperature up to $120\,\textrm{mK}$, and then a rapid increase due to thermally generated quasiparticles.  (b) Experimental measurement of relaxation rate $\Gamma_1$ for a phase qubit.  The relaxation decreases, and then increases with temperature, in qualitative agreement with theory.  The slow change at $T_p \gtrsim 140\,\textrm{mK}$ is probably due to non-idealities, such as the small occupation of the ground state (greater than $90\%$ for $T_p<100\,\textrm{mK}$, but only $77\%$ at $T_p=130\,\textrm{mK}$ and $55\%$ at $T_p=160\,\textrm{mK}$).  }
\label{figQPT}
\end{figure}

The calculated temperature dependence of the decay rate is plotted in Fig.\,\ref{figQPT}(a), which shows negligible change at low temperatures $T \lesssim 40\,\textrm{mK}$, and then a small \textit{reduction} in rate up to $T_p= 120\,\textrm{mK}$.  This unusual prediction of an improvement in decay rate with increasing temperature results from non-equilibrium quasiparticles having, on average, larger energy and thus smaller contributions from density of states and coherence factors.  At higher temperatures, the thermal generation of quasiparticles rapidly increases the decay rate.  This temperature dependence well explains data taken on superconducting resonators\cite{Barends09}.  Measurements of $\Gamma_1(T)$ for a phase qubit\cite{UCSBphaseT1} is shown in Fig.\,\ref{figQPT}(b), where a reduction and then an increase in decay rate is in qualitative agreement with predictions\cite{Fdependence}.  The simplicity of our model, which does not account for the small occupation of the qubit ground state, diffusion of quasiparticles, or gap inhomogeneities, may explain the differences.

\begin{figure}[t]
\includegraphics[width=0.47\textwidth,clip=True]{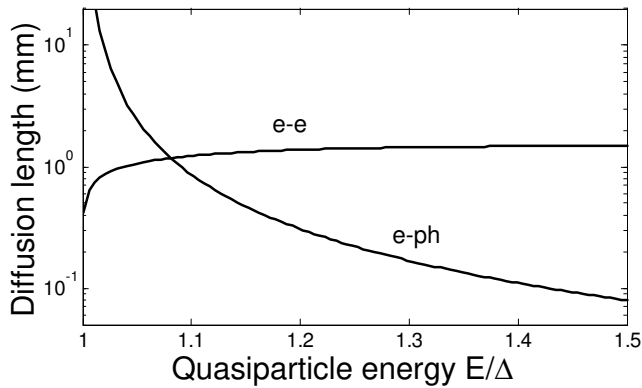}
\caption{Plot of average quasiparticle diffusion length versus energy for both the electron-phonon $1/\Gamma^s$ (e-ph) and electron recombination $1/\Gamma^r$ (e-e) decay times.  The e-e interaction dominates for $E \lesssim 1.1\,\Delta$.  Parameters are $D= 60\,v_\textrm{qp}\,\textrm{cm}^2/\textrm{s}$ (Al), $T_p=0$, and $n_\textrm{qp}=10/\mu\textrm{m}^3$. }
\label{figDlength}
\end{figure}

Previous experiments have used engineered gap\cite{gap} and trap\cite{Joyez1994,trap} structures to reduce the density of quasiparticles.  Are they large enough to be completely effective?  The quasiparticle diffusion constant for aluminum is $D= 60\,v_\textrm{qp}\,\textrm{cm}^2/\textrm{s}$, where $v_\textrm{qp}=(1-\Delta^2/E^2)^{1/2}$ is the normalized velocity, which depends on energy and approaches zero at the gap.  Quasiparticles diffuse a distance $(D\tau)^{1/2}$ in time $\tau$; we plot this length in Fig.\,\ref{figDlength} for both the electron-phonon and electron-electron decay times.  We estimate that the characteristic distance for quasiparticles to equilibrate are of millimeter lengths, much larger than the size of qubit devices.  In normal metal traps, quasiparticles lose energy at much short length scales, of order $30\,\mu\textrm{m}$.  Finally, quasiparticle energy is ultimately removed from the superconductor by recombination and emission of a phonon with energy $\gtrsim 2\Delta$.  Because phonons are long-lived at low temperature, especially for crystalline substrates, they may ballistically travel across the chip and have their energy redeposited as quasiparticles anywhere in the superconductor.  All of these estimates imply that a conservative design should have quasiparticle trapping structures of millimeter or greater size, not just local structures placed around the tunnel junctions.

Although the density of quasiparticles depends on experimental details of chip design, it is possible to make rough power estimates to help understand the generation mechanism.  The model presented here found a quasiparticle injection rate $r_\textrm{qp} \sim 2.4\cdot10^3/\textrm{s}\,\mu\textrm{m}^3$.  Assuming a total superconductor volume of $0.1\,\mu\textrm{m} \times 10\,\textrm{mm}^2$ and an energy $\Delta$ per quasiparticle, the total power to the chip is $0.06\,\textrm{pW}$.

The power load from various sources can be compared with this number.  Cosmic rays have a flux of $\sim 0.6/\textrm{cm}^2\textrm{s}$ and deposit an energy of $\sim 1\,\textrm{MeV/mm}$, yielding for a $50\,\textrm{mm}^2$ chip of thickness $0.5\,\textrm{mm}$ a power $0.02\,\textrm{pW}$.  Background radioactivity is typically of order that coming from cosmic rays.  If coaxial lines allows $\sim 4\,\textrm{K}$ thermal radiation to be transmitted with a bandwidth $100\,\textrm{GHz}$, a power $\sim 5\,\textrm{pW}$ is found.  Blackbody radiation from $1\,\textrm{K}$ gives about $60\,\textrm{pW}$ of power load to the outside of the chip mount.  Materials also slowly release energy at low temperatures, with amorphous $\textrm{SiO}_2$, polycrystalline Al, and teflon showing heat release of $\sim 100\,\textrm{pW/gr}$, $\sim 20\,\textrm{pW/gr}$, and $\sim 2\,\textrm{pW/gr}$ respectively that slowly decays with time\cite{TLS}.  From these estimates, non-equilibrium quasiparticles are clearly plausible.

In conclusion, we have developed a model that predicts qubit energy decay from the density of non-equilibrium quasiparticles.  Predictions of the model are in reasonable agreement based on a prior experimental measurement and, at a minimum, show that this decoherence mechanism should be carefully considered.   The quasiparticle energy distribution calculated from simple electron-phonon interaction qualitatively agrees with measurements, and we argue that engineered gap and trap structures should be of large (millimeter) size.  With a good understanding of the basic physics, we believe experiments can be performed to more carefully test for non-equilibrium quasiparticles, and to ultimately improve the coherence of superconducting qubits.

We thank M. Devoret, P. Delsing, and B. Mazin for helpful discussions.  This work was supported by IARPA under grant W911NF-04-1-0204 and by the NSF under grant CCF-0507227.


\begin{thebibliography}{99}
\bibitem{noteHO}
For simplicity, we have assumed matrix elements for the qubit given by the harmonic oscillator.
\bibitem{noteequiv}
The small difference in formulas arise from our consideration of only energy-decay mechanisms, whereas the Mattis-Bardeen formula accounts for both loss and gain.
\bibitem{Fdependence}
We observed no dependence of $\Gamma_1$ on $E_{10}/h$ for a $\sim\,15\%$ change in frequency from the current bias.
\end{thebibliography}
\end{document}